\documentclass[conference]{IEEEtran}
\IEEEoverridecommandlockouts
\usepackage{cite}
\usepackage{amsmath,amssymb,amsfonts}
\usepackage{algorithmic}
\usepackage{graphicx}
\usepackage{textcomp}
\usepackage{xcolor}

\usepackage[utf8]{inputenc}
\usepackage{amssymb}
\usepackage{graphicx}
\usepackage{amsmath}
\usepackage[normalem]{ulem}

\usepackage{amsfonts}%
\usepackage{amssymb}%
\usepackage{graphicx}
\usepackage{url}
\usepackage{booktabs}
\usepackage{comment}
\usepackage{bbold}
\usepackage{placeins}
\usepackage{pdfpages}
\usepackage{adjustbox}
\usepackage{enumerate}
\usepackage{mathtools}
\usepackage{longtable}
\usepackage{blkarray}
\usepackage{subcaption}
\usepackage{soul}
\usepackage[margin=1.25in]{geometry}

\usepackage{fancyhdr}

\usepackage{tikz}  
  
\usetikzlibrary{shapes, arrows, calc, arrows.meta, fit, positioning} % these are the parameters passed to the library to create the node graphs  
\tikzset{  
    -Latex,auto,node distance =1.5 cm and 1.3 cm, thick,% node distance is the distance between one node to other, where 1.5cm is the length of the edge between the nodes  
    state/.style ={ellipse, draw, minimum width = 0.9 cm}, % the minimum width is the width of the ellipse, which is the size of the shape of vertex in the node graph  
        state2/.style ={rectangle,rounded corners,dashed, draw, text width=1.7cm ,align=center}, % the minimum width is the width of the ellipse, which is the size of the shape of vertex in the node graph  
    point/.style = {circle, draw, inner sep=0.18cm, fill, node contents={}},  
    bidirected/.style={Latex-Latex,dashed}, % it is the edge having two directions  
    el/.style = {inner sep=2.5pt, align=right, sloped}  
}

\newcommand{\bb}[1]{\boldsymbol{#1}}

\newcommand{\stanx}{\boldsymbol{\mathrm{x}}}

\makeatletter

\def\ps@IEEEtitlepagestyle{
  \def\@oddfoot{\mycopyrightnotice}
  \def\@evenfoot{}
}

\def\mycopyrightnotice{

  {\footnotesize 978-1-6654-1211-7/22/\$31.00~\copyright~2022 IEEE\hfill} % <--- Change here
  \gdef\mycopyrightnotice{}

}

\newcommand{\linebreakand}{%
  \end{@IEEEauthorhalign}
  \hfill\mbox{}\par
  \mbox{}\hfill\begin{@IEEEauthorhalign}
}

\makeatother

\begin{document}

\title{Propagating uncertainty in a network of energy models  
\thanks{This work was funded by the EPSRC National Centre for Energy Systems Integration through the flex fund award FFC3-008, and by the Alan Turing Institute grant `Managing Uncertainty in Government Modelling'. VV, NS and HPW were contracted by the Turing Institute when this work was carried out, and JQS, PGC and CJD are Turing Fellows at the Institute.}
}

\author{\IEEEauthorblockN{Victoria Volodina}
\IEEEauthorblockA{\textit{Clinical Operational}\\ \textit{Research Unit}\\ 
\textit{University College London}\\
London, UK \\
v.volodina@ucl.ac.uk}
\and
\IEEEauthorblockN{Nikki Sonenberg}
\IEEEauthorblockA{\textit{Heilbronn Institute of}\\ \textit{Mathematical Research} \\
\textit{University of Bristol}\\
Bristol, UK \\
nikki.sonenberg@bristol.ac.uk}
\and
\IEEEauthorblockN{Jim Q. Smith}
\IEEEauthorblockA{\textit{Department of Statistics} \\
\textit{Warwick University}\\
Coventry, UK \\
j.q.smith@warwick.ac.uk}\\ 
%\and
\linebreakand
\IEEEauthorblockN{Peter G. Challenor}
\IEEEauthorblockA{\textit{Department of Mathematics} \\
\textit{University of Exeter}\\
Exeter, UK \\
p.g.challenor@exeter.ac.uk}
\and
\IEEEauthorblockN{Chris J. Dent}
\IEEEauthorblockA{\textit{School of Mathematics} \\
\textit{University of Edinburgh}\\
Edinburgh, UK \\
chris.dent@ed.ac.uk}
\and
\IEEEauthorblockN{Henry P. Wynn}
\IEEEauthorblockA{\textit{Department of Statistics} \\
\textit{London School of Economics}\\
London, UK \\
h.wynn@lse.ac.uk}
}

\maketitle

\begin{abstract}
  
Computer models are widely used in decision support for energy systems operation, planning and policy. A system of models is often employed, where model inputs themselves arise from other computer models, with each model being developed by different teams of experts. Gaussian Process emulators can be used to approximate the behaviour of complex, computationally intensive models and used to generate predictions together with a measure of uncertainty about the predicted model output.
This paper presents a computationally efficient framework for propagating uncertainty within a network of models  with high-dimensional outputs  used for energy planning. We present a case study from a UK county council 
considering low carbon technologies to transform its infrastructure to reach a net-zero carbon target. The system model considered for this case study is simple, however the framework can be applied to larger networks of more complex models.

\end{abstract}


\section{Introduction}
Computer models are widely used in decision support for energy  systems operation, planning and policy.  A  network of models is often employed, where model inputs themselves arise from other computer models,  with each model being developed by different experts from across various disciplines. To perform inferences on a single computer model such as calibration, prediction, uncertainty and sensitivity analysis, we require multiple simulation runs.
However, as the computer model can be very expensive to operate, we need to acknowledge the uncertainty about its outputs at unseen input parameter values, namely code uncertainty \cite{Kennedy2001}. This is relevant for the analysis of networks of models, since the outputs together with uncertainties from the first layer computer models determine the uncertain inputs in the second layer computer models and so on.  
For principled decision support under UK government guidelines \cite{Treasury2015}, the uncertainties associated with individual models' outputs need to be explicitly quantified and propagated across the network. Other related frameworks for the coupling of probabilistic systems include \cite{idss2015}.

Gaussian Process (GP) emulators are commonly used to approximate the behaviour of complex, computationally intensive  computer models; an emulator provides predictions and quantifies uncertainty in the state of knowledge regarding the behaviour of the model including that arising from only a limited number of runs being possible. 
  GP emulators have been widely used as \emph{surrogates} to complex computer models in climate and environmental studies \cite {Conti2009,volodina2020diagnostics,williamson2014evolving} and electricity prices \cite{Wilson2018}.   
  
This paper presents a framework for propagating uncertainty within a network of models with high-dimensional outputs (time-series) using recent methodological advances in the theory of networks of GP emulators \cite{Ming2021,Sanson2019,Kyzyurova2018}. First, we employ  principal component analysis (PCA) to reduce the dimension of the output space  by projecting the high-dimensional output onto a low-dimensional basis, we then specify a univariate GP emulator for the individual coefficients of a low-dimensional basis. The efficient propagation of uncertainty between models is achieved by passing the first and second moments between the probability models for the basis vectors' coefficients obtained and projecting back to reconstruct the mean and variance for the original model output.

This methodology is applied to a case study for a UK county council considering low carbon technology options to transform the infrastructure at a facility in order to reach a carbon zero target in 2050 \cite{BEIS2020}. Under the Public Decarbonisation Scheme \cite{DecarbScheme}, the council plans to replace the gas boiler with an electric-powered ground source heat pump (GSHP) to supply heating.  The simple network model is used to demonstrate how projections of operational costs of the  gas boiler can be generated, however the framework is scalable and can be applied to larger networks of more complex models.
  
In Section \ref{sec:methodology} we describe the GP methodology, and  in Section \ref{sec:energysystems} we describe the individual computer models used in energy systems planning. We present results for the case study in Section \ref{sec:results} and in Section \ref{sec:discussion} we provide a discussion.

\section{Methodology}\label{sec:methodology}

In Sections \ref{sec:GP_model} \& \ref{subsec:multiGP}  we define the univariate GP model and  multivariate GP models and use these to assemble the linked GP emulator in Section \ref{subsec:prob_int}.
 
\subsection{Gaussian Process model}
\label{sec:GP_model}
%Here we describe the Gaussian Process models for individual components, that will then be coupled together to propagate uncertainty within a system of models.  
Let $\stanx=(x_1, \dots, x_p)\in\mathbb{R}^p$ be a $p$-dimensional vector of inputs and $f(\stanx)$ be the scalar-valued output that represents the process of interest. The GPs are fully specified by mean function $\mu(\cdot)$ and covariance function $\sigma^2r(\cdot, \cdot; \bb{\delta})$, where $\bb{\delta}$ and $\sigma^2$ correspond to the vector of correlation length parameters and variance parameter respectively.

%We start by defining a statistical model for $f(\stanx)$ as a sum of three processes
%\begin{equation}
%    f(\stanx)=\bb{h}(\stanx)^T\bb{\beta}+\epsilon(\stanx)+\nu(\stanx),
%\end{equation}
%where $\bb{h}(\stanx)^T\bb{\beta}$ represents a global response surface behaviour, $\epsilon(\stanx)$ is a correlated residual process capturing local input dependent deviation from the global response surface (modelled as a zero-mean Gaussian process with covariance function $k(\stanx, \stanx'; \sigma^2, \bb{\delta})=\sigma^2r(\stanx, \stanx'; \bb{\delta})$), and $\nu(\stanx)$ is a nugget process representing the noise in the response (modelled as a zero-mean Normal with variance $\tau^2$). 

Suppose we observe $n$ realisations $F^{\mathcal{T}}=(f(\stanx_1), \dots, f(\stanx_n))$ at design points $X^{\mathcal{T}}=(\stanx_1, \dots, \stanx_n)$.
The emulator is fitted based on an ensemble of runs of the model $f$, denoted by $\mathcal{D}=\{X^{\mathcal{T}}, F^{\mathcal{T}} \}$, using a Bayesian approach with a non-informative prior for parameters $\sigma^2$ and $\bb{\delta}$ \cite{Haylock1996}.

\subsection{Multivariate Gaussian Process model}
\label{subsec:multiGP}
For computer models with multivariate outputs given by an $l$-dimensional vector $\bb{f}(\stanx)=(f_1(\stanx), \dots, f_l(\stanx))$,
for example, time series output, the data can be projected onto a low-dimensional basis using principal components \cite{Higdon2008}, and then  independent GP models are specified for coefficients of this basis.

The $n$ computer model simulations are stored in an $l\times n$ matrix $\bb{F}=(\bb{f}(\stanx_1), \dots, \bb{f}(\stanx_n))$. After subtracting out the mean simulation, $\bb{\mu}$, and scaling the simulation output to obtain the centred ensemble, $\bb{F}_{\mu}$, the singular value decomposition for the centred ensemble can be written as
\begin{equation}
    \bb{F}_{\mu}^T=\bb{U}\bb{\Sigma}\bb{V}^T.
\end{equation}
We then obtain the principal component basis, denoted by $\bb{\Gamma}=(\bb{\gamma}_1, \dots, \bb{\gamma}_{n-1})$, which are the first $n-1$ columns of $\bb{V}$ and where each individual basis vector $\bb{\gamma}_i$ has length $l$.

Given this basis, $\bb{f}(\cdot)$ can be written as a linear combination of the basis vectors:
\begin{align}
\label{eq:SVDf}
    \bb{f}(\stanx)-\bb{\mu}&=\sum_{i=1}^{n-1}\bb{\gamma}_ic_i(\stanx)+\bb{\epsilon},\\
    &=\bb{\Gamma}\bb{c}(\stanx)+\bb{\epsilon},
\end{align}
where $c_i(\stanx)$ is the coefficient for basis vector $\bb{\gamma}_i$ and $\bb{\epsilon}$ is a residual vector.

Exploiting the fact that the basis vectors of $\bb{\Gamma}$ are orthogonal, we can fit univariate Gaussian process emulators for the coefficients $c_i(\cdot)$ for each basis vector separately. We choose the first $q$ vectors and denote the truncated basis by $\bb{\Gamma}_q = (\bb{\gamma}_1, \dots, \bb{\gamma}_q)$.
%\begin{equation}
%\bb{\Gamma}_q = (\bb{\gamma}_1, \dots, \bb{\gamma}_q).
%\end{equation}
%We emulate the coefficients for the first $q$ basis vectors, then conditioned on the projected ensemble, we can obtain the posterior distribution for the reconstruction of $\bb{f}(\cdot)$ \cite{Higdon2008}.
Conditioned on the projected ensemble, we emulate the coefficients for the  $q$ basis vectors and obtain the posterior distribution for the reconstruction of $\bb{f}(\cdot)$.

\subsection{Linked Gaussian Process emulators}
\label{subsec:prob_int}
To perform the coupling of a network of feed-forward computer models, we present the linked GP emulator methodology used for uncertainty quantification. Consider a simple system, where a $d$-dimensional output produced by computer models in the first layer feed into a computer model in the second layer \cite{Ming2021}.

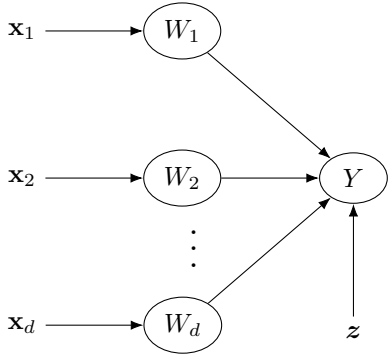
\begin{figure}[h!]
\begin{center}
\begin{tikzpicture}  
    \node[draw=none,fill=none] (a) at (0,0) {  $\stanx_1$};  
      \node[draw=none,fill=none] (c) [below =of a] {  $\stanx_2$};  
        \node[state] (b) [right =of a] { $W_1$}; 
            \node[state] (d) [right =of c] { $W_2$}; 
            \node[state] (e) [right =of d] { $Y$};  
              \node[draw=none,fill=none] (f) [below =of c] {  $\stanx_{d}$};  

            \node[state] (g) [right =of f] { $W_{d}$}; 

              \node[draw=none,fill=none] (h) [below =of e] {  $\bb{z}$};  
     \path (a) edge (b);
    \path (b) edge (e);  
    \path (c) edge (d);  
    \path (d) edge (e);  
    \path (f) edge (g);  
    \path (g) edge (e);  
    \path (h) edge (e); 
    \path (d) -- (g) node [font=\Large, midway, sloped] {$\dots$};
 
\end{tikzpicture} 
\end{center}
\caption{A feed-forward graph depicting the relationship between random variables $W_1, W_2, \dots, W_d$ and $Y$.
%are univariate GP emulators.
} 
\label{fig:LinkedGPs}
\end{figure}

For $i=1, \dots, d$ define a GP emulator to approximate the outputs produced by individual computer models
  \begin{align}
        W_i\vert \stanx_i\sim \text{GP}(\mu_i(\cdot), \sigma_i^2 r(\cdot, \cdot; \bb{\delta}_i)),
    \end{align}
  and 
  \begin{align} Y\vert W_1, \dots, W_d, \bb{z}\sim \text{GP}(\mu_y(\cdot), \sigma_y^2r(\cdot, \cdot; \bb{\delta}_y)),
      \end{align}
\begin{comment}
  \begin{align}
        w_i\vert \stanx_i\sim \text{GP}(\mu_i(\cdot), \sigma_i^2 r(\cdot, \cdot; \bb{\delta}_i)),
    \end{align}
  and 
  \begin{align} y\vert w_1, \dots, w_d, \bb{z}\sim \text{GP}(\mu_y(\cdot), \sigma_y^2r(\cdot, \cdot; \bb{\delta}_y)),
      \end{align}
\end{comment}
where $\stanx_1, \dots, \stanx_d$ and $\bb{z}$ are vectors of input parameters. The outputs of the computer models are treated as random variables, and the relationships between these are shown in Figure~\ref{fig:LinkedGPs}. 

The  distribution of $Y\vert \stanx_1, \dots, \stanx_d, \bb{z}$ can be written as, 
\begin{align*}
\label{eq:integral}
    p(y| \stanx_1,.., \stanx_d, \bb{z})=\int p(y |\bb{w}, \bb{z})p(\bb{w}| \stanx_1,.., \stanx_d)d\bb{w},
\end{align*}
where $\bb{w}=(w_1, \dots, w_d)^T$. However, $(y\vert \stanx_1, \dots, \stanx_d, \bb{z})$ is neither analytically tractable nor Gaussian in general.

The first two moments can be computed by Monte Carlo samples under the assumption that the densities inside the integral are Gaussian \cite{Sanson2019}. Further, under some mild conditions, the first two moments can be calculated analytically \cite{Ming2021,Kyzyurova2018}. 
In particular, suppose we observe realisations of $y$ and $\bb{w}$, denoted by $Y^{\mathcal{T}}$ and $W^{\mathcal{T}}$, at design (training) sets $X^{\mathcal{T}}$ and $Z^{\mathcal{T}}$, let $\mathcal{D}=\{X^{\mathcal{T}}, Z^{\mathcal{T}}, W^{\mathcal{T}}, Y^{\mathcal{T}} \}$. For specific classes of correlation functions, \cite{Ming2021} presented the closed-form expressions for the mean and variance of $y$ at new inputs $\tilde{\stanx}$ and $\tilde{\bb{z}}$, conditioned on $\mathcal{D}$. 

We fit univariate GP emulators for the coefficients of the retained basis vectors obtained for multivariate outputs of component models, and substitute these parameter estimates in the closed form expression of \cite[Theorem 3.1]{Ming2021}. Conditioned on the projected ensemble, we obtain a posterior mean and variance for the output of the composite model for a given input.
%For practical applications as in this paper, we fit GP emulators for each component model individually, and then substitute these parameter estimates in the closed form expression \cite[Theorem 3.1]{Ming2021} to obtain a  prediction and a measure of uncertainty for the output of the composite model for a given input. %\cjd{Need some comment as to why the required conditions are valud here.}
%\textcolor{blue}
%We fit univariate GP emulators for the coefficients of the retained basis vectors, since the basis vectors are orthogonal.

 \section{ Component models}\label{sec:energysystems}

%This section describes  the component models of the decision support framework and are used to generate operational costs and carbon emissions from 2020-50 to support the planning decision for the Leisure Centre case study. We provide details for each of the two models, followed by their corresponding emulator in Section \ref{sec:emuated_components}.

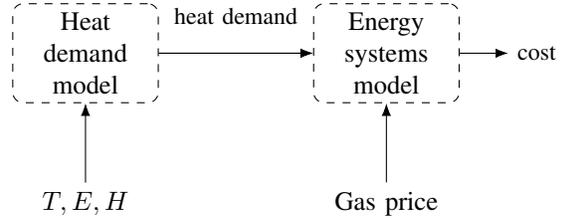
\begin{figure}[h!]
\begin{center}
\begin{tikzpicture}  
    \node[draw=none,fill=none] (a) at (0,0) {  $T, E, H$}; 
    \node[state2] (b) at (0,2) {  Heat demand model};  
    \node[state2] (c) at (4,2) { Energy systems model};  
        \node[draw=none,fill=none] (d) at (4,0) { Gas price};
     \node[draw=none,fill=none] (e) at (2, 2.5) {\small heat demand};
     \node[draw=none,fill=none] (f)
     at (6, 2) {\small cost};
    \path (a) edge (b);
    \path (b) edge (c);  
 
     \draw (b) -- (c);  
     \draw (d) -- (c);  
     \draw (c) -- (f);
\end{tikzpicture} 
\end{center}
\caption{A network of models for energy planning, where $T$ is the surface temperature, $E$ is the efficiency of the equipment and $H$ is the global building transmission coefficient.}

%\textcolor{green}{[Do we need the "heat demand" label on the arrow from Y1 to Y2 - seems redundant]}
 
\label{fig:ModelDes}
\end{figure}

In this section we  provide details for each of the component models of the decision support framework illustrated in Figure \ref{fig:ModelDes} for the case study from a UK county council 
considering low carbon technologies to transform its infrastructure to reach a net-zero carbon target.

\begin{comment} 
\begin{figure}[h!]
\begin{center}
\begin{tikzpicture}  
    \node[draw=none,fill=none] (a) at (0,0) {  $\stanx_1$}; 
    \node[state2] (b) at (2,0) {  Heat demand model $Y_1$ };  
    \node[state2] (c) at (4.5,0) { Energy systems model $Y_2$ };  
        \node[draw=none,fill=none] (d) at (4.5,1.5) { $\stanx_2$}; 

    \path (a) edge (b);
    \path (b) edge (c);  
 
     \draw (b) -- (c);  
     \draw (d) -- (c);  
\end{tikzpicture} 
\end{center}
\caption{\textcolor{green}{A feed-forward graph with vector of inputs $\stanx_1$  (surface temperature, efficiency of the equipment, global building transmission coefficient) to the heat demand model with output $Y_1$, and with gas price input  $\stanx_2$ to the \textcolor{red}{energy} systems model with  output  $Y_2$,  operational cost .}}
 
\label{fig:ModelDes}
\end{figure}
\end{comment}

\subsubsection{Heat demand model}\label{subsec:hddmodel}
\label{sec:energy_demand_model}

Heat demand is calculated based on the degree days statistic (i.e., the sum over days in which temperature is below a given temperature threshold  \cite{DeRosa2014,Larsen2020,Spinoni2018}). We vary the three inputs: surface temperature $(T)$, efficiency of the equipment $(E)$ and the global building transmission coefficient $(H)$ (which when multiplied by the degree days gives the heating power required). Surface temperature takes the form  of a time series of annual averages from 2021 to 2050, whereas the other inputs are assumed to be fixed throughout the whole period. Projections of the annual heating demand are then generated up to 2050. For details on this model we refer to \cite{DeRosa2014}. 

We derive the domains for the input parameters 
from the half-hourly energy data for 2017-2021 provided by the council together with the historical surface temperature data for 2017-2020.

\subsubsection{Energy systems model}
 
This model is based on the widely used OSeMOSYS open source framework \cite{howells2011}, that computes the energy supply mix (in terms of generation capacity and energy delivery) given the heat demand, fuel prices, and the technologies connected to the system. The model considers representative days for each of the four seasons, and daily/seasonal storage technologies which shift demand from day to night and between seasons, respectively. Based on the half-hourly demand data provided by the local council, the share of annual heat demand attributed to each season and day/night is displayed in Table \ref{Distribution_HeatDemandReport} and specified in the energy systems model.

\begin{scriptsize}
    \begin{table}[h!]
    \begin{center}
    \caption{Share of annual heat demand (\%)}
     \label{Distribution_HeatDemandReport}
         \begin{tabular}{lllll}
            \toprule
          & Winter&Spring&Summer&Autumn\\
            \midrule   
              Day & 26.5& 17.7&12.2&24.5\\
              Night & 4.66&5.11&4.12&5.14 \\
         \bottomrule
        \end{tabular}
            \end{center}
    \end{table}    

\end{scriptsize}
 For this case study, we are interested in output produced by the model of the total operational cost, a time-series from 2021 to 2050. The model inputs considered are heating demand and gas price, that take the form of an annual average time-series from 2021 to 2050 \cite{BEIS2020}.

\section{Results}
\label{sec:results}

\subsection{Emulation}\label{sec:emuated_components}
To construct the decision support system, we first build the GP emulators to approximate outputs produced by the individual component models in Sections \ref{sec:GP_demand} and  \ref{sec:GP_heat_model}. In Section \ref{sec:linkedgpmodels}, we outline the process of linking the GP emulators with mutlivariate output.

\subsubsection{GP emulator for the heat demand model output}
\label{sec:GP_demand}
 
The surface temperature input is represented by one parameter, a shift away from the central projection on a continuum, where a shift of $\pm 1$ corresponds to the high/low scenario, with intermediate values interpolated between these \cite{Wilson2018}.

\begin{table}[h!]
\caption{Domain of input parameters for  GP emulators for the heat demand model output.}
\centering
\begin{tabular}{l c} 
 \toprule
 \textbf{Input parameter (unit)} & \textbf{Domain} \\
 \midrule
Shift parameter in surface temperature & $[-1, 1]$\\
 Efficiency of the equipment $(E)$ & $[0.5, 1]$\\
 Global building transmission coefficient $(H)$ & $[5, 20]$\\
 \bottomrule
\end{tabular}
\label{table:domain_demand}
\end{table}

Table \ref{table:domain_demand} provides the domains of the input parameters considered for emulation. We use the maximin distance Latin Hypercube (LHC) to generate a space-filling design to explore the output behaviour across the input space and fit the emulator \cite{Morris1995}. The output of interest is a vector of annual heating demands from 2021 to 2050, which is projected onto a low-dimensional basis by PCA. The first two principal components were found to explain 98\% of the total variance and are retained for our analysis.

We consider coefficients of this basis, denoted by $c_i^{(1)}(\stanx_1)$, $i=1, 2$, as  functions of the input parameters and construct a GP emulator for each coefficient as in  \ref{sec:GP_model}. The regression function $h(\stanx_1)=(1, x_{11}, x_{12}, x_{13})$ includes a constant, and linear terms in each component of $\stanx_1$ \cite{Kennedy2001}. The correlation function in the GP is a squared exponential, and for the fit the \texttt{RobustGaSP} package  \cite{RobustGP2020} is used. 

\begin{figure}[h!]
\begin{center}
\includegraphics[width = 1\columnwidth]{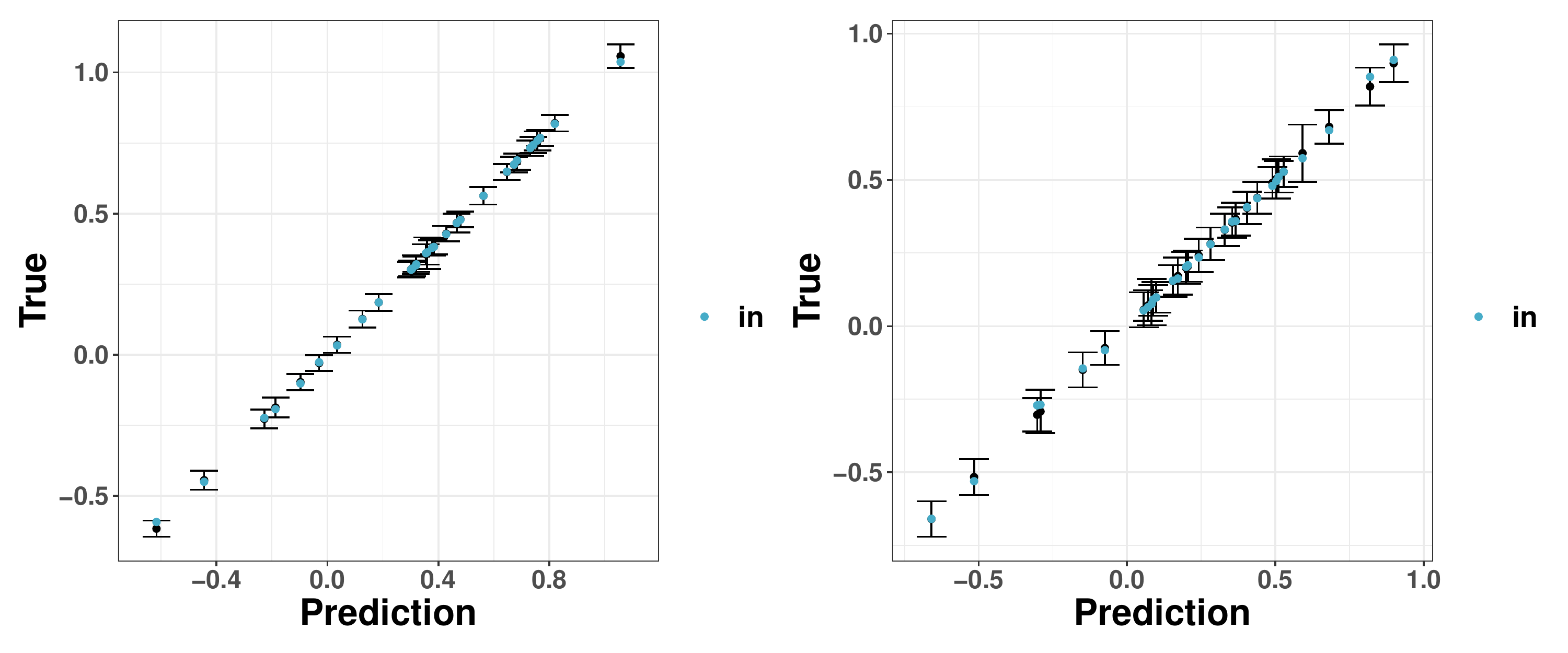}
\includegraphics[width = 1\columnwidth]{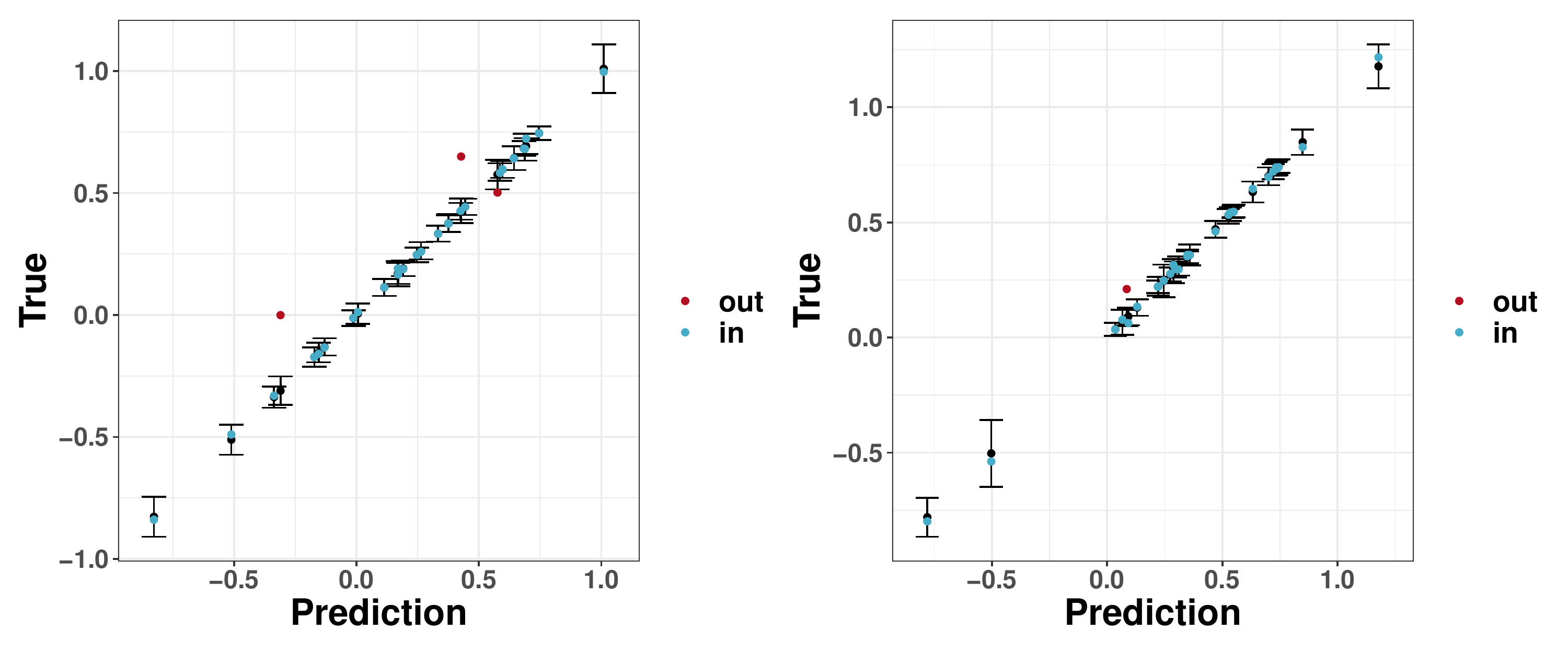}
\end{center} 
\caption{Cross-validation plots for emulators  for the coefficients of the two basis vectors for the heat demand model
(\textit{top}) and energy systems model (\textit{bottom}).} 
\label{fig:PCALOO}
\end{figure}

To validate the performance of the GP emulators,  a test set of size 30 is used. Figure~\ref{fig:PCALOO} (top row) presents the cross-validation diagnostics for each basis coefficient.  We plot the emulated values and the model outputs on the x-axis and y-axis respectively.  The black points and error bars represent the emulator prediction and a two standard deviation prediction interval. The true model values are green if they lie within two standard deviation prediction intervals, or red otherwise.   
We observe that the emulator predictions lie close to the true values, and the size of the error bars is small for both basis coefficients, indicating that the emulator is performing well.

\subsubsection{GP emulator for energy systems model output}
\label{sec:GP_heat_model}

Similarly, we construct an emulator  to approximate the output of the energy systems model. The heating demand input is replaced by  the two basis coefficients obtained in  \ref{sec:GP_demand}, whereas the gas price input is replaced by a shift parameter, with high, low and baseline  scenarios again represented by values $\pm 1$ and $0$  of a shift parameter. 
The output of interest is the annual operating cost from
2021-50, which is projected on to principal components, here we use the first two principal components that explain 93\% of the total variance. 

The same approach to fitting GP emulators for the retained basis coefficients, $c_i^{(2)}(\cdot), i=1, 2$, as a function of two basis coefficients and a shift parameter for gas price is followed in Figure~\ref{fig:PCALOO} (bottom row).  We observe that three of the tested model outputs were outside the prediction intervals for our coefficients of principal components, which is still consistent with our uncertainty specification.

\subsection{Linking the GP models}\label{sec:linkedgpmodels}
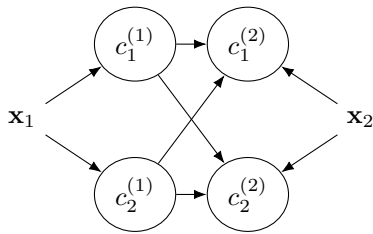
\begin{figure}[h!]
\begin{center}
\begin{tikzpicture}  
    \node[draw=none,fill=none] (a) at (-.5,0) {  $\stanx_1$}; 
     
    \node[state] (b) at (1, 1) { $c^{(1)}_1$};  
   % \node[state] (d) [below =of a] { $Y_1$};  

    \node[state] (c) at (1, -1) {$c^{(1)}_2$};  

    \node[state] (d) at (2.5, 1) { $c^{(2)}_1$};  

    \node[state] (e) at (2.5, -1) {$c^{(2)}_2$};

    \node[draw=none,fill=none] (g) at (4, 0) {$\stanx_2$};  

    \path (a) edge  (b); 
    \path (a) edge (c);  
    \path (c) edge  (d); 
    \path (b) edge  (d); 
    \path (b) edge  (e); 
    \path (c) edge  (e); 
    \path (g) edge  (d); 
    \path (g) edge  (e); 

\end{tikzpicture} 
\end{center}
     \caption{The coefficients for the linked emulators.} 
    \label{fig:ProbModelDes}
\end{figure}

The two emulators for heat demand and the energy systems model are coupled using the direct calculation method described in Section \ref{subsec:prob_int}. For the graphical model in Figure~\ref{fig:ModelDes}, in Figure~\ref{fig:ProbModelDes} we llustrate the relationship between the first two coefficients  of PCA basis vectors obtained from the heat demand model ensemble, $c_1^{(1)}$ and $c_2^{(1)}$, and the first two coefficients of PCA basis vectors obtained from the energy systems model ensemble,  $c_1^{(2)}$ and $c_2^{(2)}$, 
given the vectors of global inputs  $\stanx_1$ and $\stanx_2$.

\begin{figure}[h!]
\begin{center}
\includegraphics[width =1\columnwidth]{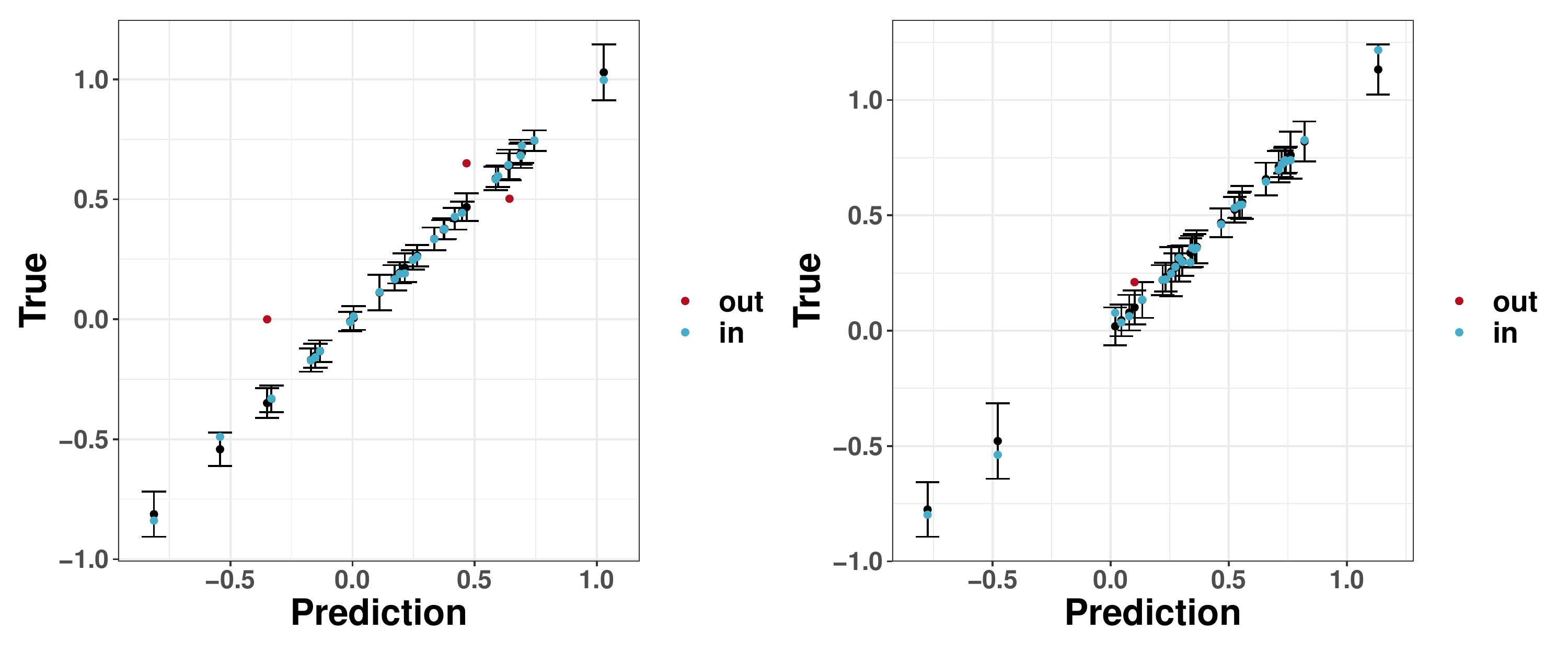}
\caption{Cross-validation plots for the linked emulators.
}\label{fig:ValidLinked}
\end{center} 
\end{figure}

 Figure \ref{fig:ValidLinked} presents the cross-validation diagnostics for the linked emulators. We continue operating with GP emulators for PCA coefficients described in  \ref{sec:GP_demand} and \ref{sec:GP_heat_model}.

\begin{figure}[h!]
\begin{center}
\includegraphics[width =1\columnwidth]{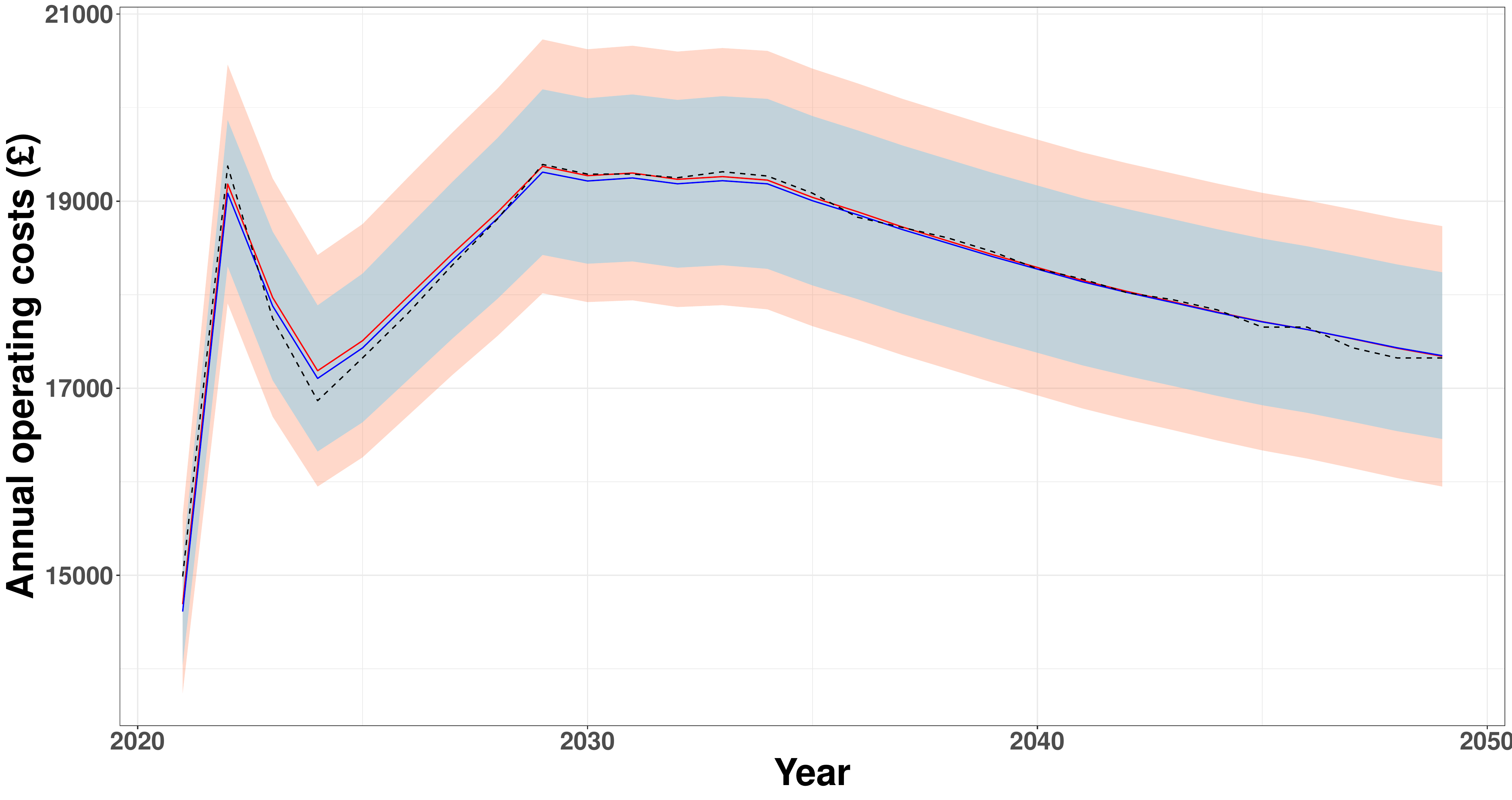}
\end{center} 
\caption{
A simulation run from the energy systems model
(\textit{dashed line}), projected mean $\pm 2\sigma$ prediction interval produced by the linked emulators (\textit{red}) and emulator for the energy systems model only (\textit{blue}).} 
\label{fig:ProjectedComparison_energy}
\end{figure}

Figure~\ref{fig:ProjectedComparison_energy} shows the projections of annual operational costs. The projected mean (solid line) and two standard deviation prediction interval (shaded region) produced by linked emulators (red); and emulation for the energy systems model only taking full runs of the demand model as inputs (blue). The dashed line is the simulation run produced by the energy systems model.

The linked emulator allows the consideration of uncertainty from both the heat demand model and energy systems model; it can be seen in the emulation of only the energy system model output, the uncertainty of the projection is underestimated compared to the more comprehensive treatment. A limited treatment of uncertainty such as the blue series can have significant consequences in practical decision situations, through consequent projections being overconfident.

\section{Discussion}\label{sec:discussion}

Simulator models are commonly used to provide decision support for those managing complex physical systems. We have demonstrated how to analyse two linked simulator models with multivariate outputs using Gaussian process emulators, a well-established approach for quantifying uncertainty in computer models \cite{Williamson2012}. The presented approach is more efficient compared to performing Monte Carlo simulation with full probability distributions for the basis vectors' coefficients.
To our knowledge, this is the first application of such an approach for decision support within the energy systems domain.

The developed method has been applied to a planning question proposed for a facility managed by a UK county council, namely a replacement of a gas boiler with a ground source heat pump. We presented projections for the total operating costs up to 2050, considering the uncertainties associated with the environment under which such new systems will need to operate. Under the carbon zero policy target the  cost projections for alternative heating facilities can be used to support the decision making process in local councils when various low carbon technologies are considered for long term planning.

The system model considered for this case study is simple, however the framework can be applied to larger networks of more complex models within energy systems planning. Decision support frameworks such as this, which provide direct  expressions for the efficient propagation of uncertainty between component models are critical for scalability.  Further work within this application domain includes extending the network model to include other variables, and studying the effect of storage on future projections as well as smoothing supply-demand fluctuations on operating timescales.

\section*{Acknowledgements}
We would like to thank Mark Roberts (Northumberland County Council) for his constructive and substantive comments.

\bibliographystyle{IEEEtran} 
\bibliography{library}

\end{document}